\documentclass[prl,twocolumn,showpacs,amssymb,floatfix]{revtex4}
\usepackage{graphicx} \usepackage{subfigure}

\begin{document}

\title{Elasticity from the Force Network Ensemble in Granular Media}

\author{Srdjan Ostojic and Debabrata Panja} \affiliation{Institute for
Theoretical Physics, Universiteit van Amsterdam, Valckenierstraat 65,
1018 XE Amsterdam, The Netherlands}

\date{\today}

\begin{abstract} 
Transmission of forces in static granular materials are studied within
the framework of the force network ensemble, by numerically evaluating
the mechanical  response of hexagonal packings  of frictionless grains
and rectangular  packings of frictional grains.  In  both cases, close
to  the  point  of  application  of  the  overload,  the  response  is
non-linear and displays two peaks, while at larger length-scales it is
linear and  elastic-like. The cross-over between  these two behaviors
occurs at a  depth that increases with the  magnitude of the overload,
and decreases with increasing friction.
\end{abstract}

\pacs{45.70.-n, 45.70Cc, 46.65+g}

\maketitle

Mechanical properties  of granular materials, or  assemblies of static
macroscopic particles  have attracted  much attention in  recent years
\cite{jaeger:rev,degennes:rev,bouch:rev,wittmer96,reydellet01,geng01:1,geng03,mueggenburg02,goldenberg05,atman05-2,ellenbroek05,gland06,otto03,snoeijer04-1,tighe05,ostojic06,snoeijer05}.
Under  external  stresses,  granular  materials present  a  solid-like
behavior, which  is due  to an intricate  network of  repulsive forces
between  particles in contact.  The discrete  nature and  high spatial
inhomogeneity of this force network  are the source of many remarkable
phenomenological properties \cite{jaeger:rev,degennes:rev}, which have
raised doubts  about the relevance of a  continuum, elastic mechanical
description   traditionally   used   in  the   engineering   community
\cite{nedderman:book}.    As  a  result,   a  number   of  alternative
descriptions  have  been  proposed  for the  propagation  of  stresses
\cite{wittmer96,bouch:rev}.

To distinguish between the various theoretical approaches, experiments
\cite{reydellet01,geng01:1,geng03,mueggenburg02,spannuth04}      and     numerics
\cite{goldenberg05,atman05-2,ellenbroek05,gland06} examined the mechanical response
of   granular   packings   \cite{bouchaud95,degennes:rev},  i.e.   the
variation  of  forces  due   to  a  small,  localized  overload.   The
experiments identified  a strong influence of  the underlying geometry
of  grains.   In a  disordered  packing  of  polydisperse grains,  the
maximal response  was found vertically below the  point of application
of  the overload,  with the  width of  the response  function linearly
increasing with depth, as predicted by isotropic elasticity theory. In
contrast,  in a  crystalline arrangement  of monodisperse  grains, the
response is  maximal along a  set of preferred  directions, resembling
the  predictions of anisotropic  elasticity \cite{otto03}.   These two
aspects   were   reconciled    by   molecular   dynamics   simulations
\cite{goldenberg05}  suggesting  that   the  anisotropic  response  in
ordered packings is only a short-range, non-linear effect, which gives
way to isotropic elastic behavior above a cross-over length scale.

Altogether,  experimental  and  numerical results  thus  unambiguously
indicate  the  elastic  nature   of  stress  propagation  in  granular
materials.  It however remains unclear how such a picture emerges from
the mesoscopic physics, and in particular from the underlying discrete
force network.  Many aspects  of this highly disordered structure have
been  successfully described  using the  so-called {\em  force network
ensemble} \cite{snoeijer04-1}.  This  statistical approach ignores the
details  of the  contact mechanics,  and takes  into account  only the
fundamental  constraint that the  forces must  balance on  each grain.
For a given geometrical arrangement, this requirement does not specify
a unique force network, but  a whole ensemble. The central idea behind
the force ensemble  approach is that generic features  of forces might
be  described by the  statistics of  this ensemble.   Remarkably, this
approach  was found to  account for  a number  of properties  of force
networks,   in  particular  the   distribution  of   force  magnitudes
\cite{snoeijer04-1,tighe05}, the geometrical  patterns of large forces
\cite{ostojic06}, and  the anisotropy  and yielding of  force networks
under shear \cite{snoeijer05}.

\begin{figure}
\begin{center}
\includegraphics[width=0.75\linewidth]{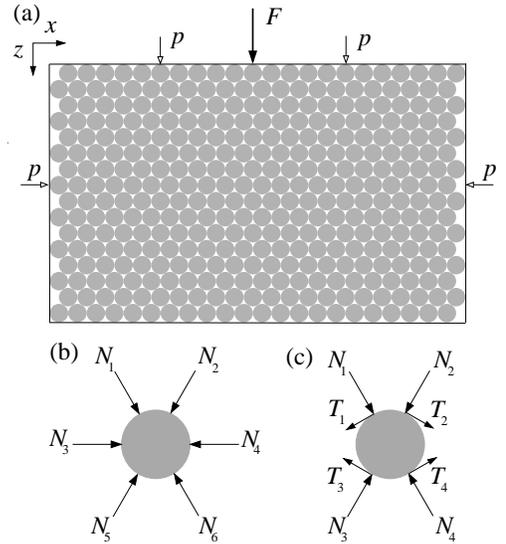}
\caption{The geometries studied: (a) ordered packing of $L$ layers 
of grains confined by a pressure $p$ on the boundaries; each layer
contains $2L+1$ grains; an overload 
$F$ is applied on the grain $x_0$ in the top layer. (b) in the
frictionless case, each grain interacts via normal forces with six
neighbors; (c) in the frictional case, each force has both a normal
and a tangential components, but there are no contacts between grains
on the same layer,
and the horizontal boundary conditions are chosen to be periodic.
\label{fig:packing}}
\end{center}
\end{figure}

\begin{figure*}
\begin{center}
\includegraphics[width=0.6\linewidth]{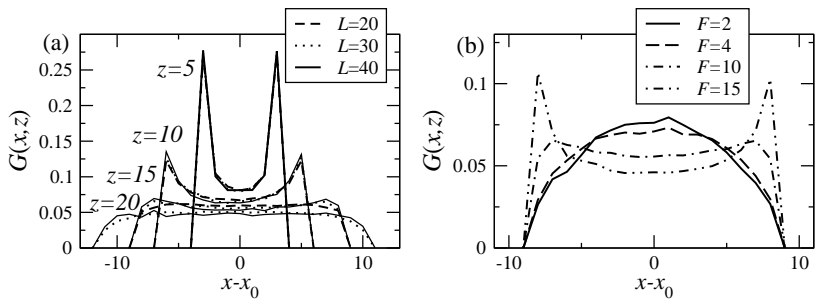}
\end{center}
\noindent \caption{Simulation results for the mean response $G(x,z)$
in the frictionless case: (a) $G(x,z)$ for $F=10$ as function of
$x-x_0$, at four different depths $z$, and for three different system
sizes $L$; (b) $G(x,z)$ at the depth $z=15$,  for four different
values of the overload 
$F$ (in units of $p$).
\label{fig:pressure}}
\end{figure*}

In this Letter, we show  that the force network ensemble describes the
transmission  of  forces  at   an  unexpected  level  of  detail.   By
numerically studying the mechanical response within this framework, we
find  a  long  length-scale  elastic  behavior as  well  as  a  short
length-scale non-linearity.   We examine two  distinct two-dimensional
settings: (i)  a hexagonal packing  of frictionless grains and  (ii) a
rectangular  packing  of  frictional   grains,  both  confined  by  an
isotropic pressure.  In both cases,  close to the point of application
of  the overload  the response  is double-peaked  and  non-linear; but
deeper into the  packing an increasing amount of  force is transmitted
towards the center of the packing.   At a given depth, a cross-over to
a single-peak  occurs.  Below that depth, the  response is essentially
linear   and  elastic-like.  The   cross-over  depth   increases  with
increasing  overload,  but,  for  frictional  grains,  decreases  with
friction.

{\it Mechanical response from the  force ensemble---} In the spirit of
Edward's  statistical approach  \cite{edwards89,bouch:rev},  the force
network ensemble ignores  the details of the dynamical  history of the
granular  packing, and considers  the ensemble  of all  possible force
networks for a given contact geometry.  This ensemble is restricted by
the constraints  that the  forces must balance  on each grain,  and be
consistent  with  imposed  boundary  stresses.   In  generic  granular
packings,  the number of  equations of  mechanical balance  is smaller
than  the   number  of  unknowns   \cite{silbert02-1}.   The  ensemble
$\cal{E}$ of allowed force  networks is thus a high-dimensional convex
set \cite{unger04:1,mcnamara04}, whose boundaries are delimited by the
requirement  that  all  normal  forces  are repulsive,  and  that  all
tangential forces  satisfy Coulomb's  condition.  By analogy  with the
microcanonical  ensemble,  every  point  of  $\cal{E}$  is  considered
equally likely.

The shape  of $\cal{E}$ intrinsically  depends on the  boundary forces
imposed  on  the  packing.   If  we call  ${\cal  E}_p$  the  ensemble
corresponding to a  uniform pressure $p$ on the  boundaries and ${\cal
E}_{p+F}$  the  ensemble  corresponding   to  pressure  $p$  with  and
additional vertical overload $F$ on a given grain of the boundary, the
average mechanical response can  be defined as $G(i,j)=(\langle W_{ij}
\rangle_{{\cal  E}_{p+F}}-\langle   W_{ij}  \rangle_{{\cal  E}_p})/F$,
where $W_{ij}$ is  the total vertical force on  the grain $(i,j)$, and
the brackets denote averages  over respective ensembles.  Note that in
the spirit of Edwards'  approach, this definition ignores the dynamics
generated by the  application of the overload, and  considers only the
possible  outcomes.    In  particular,  there  is   no  assumption  of
linearity.

To compute  $G(i,j)$, we sample  the force ensemble using  Monte Carlo
simulations.   We  first  identify  a parametrization  of  ${\cal  E}$
\cite{ostojic04:2}.   Starting   from  an  interior   point,  we  then
implement  a random walk  which is  not allowed  to leave  ${\cal E}$ \cite{vempala05}.
Such a procedure satisfies a detailed balance which leads
to a uniform sampling. 

\begin{figure*}
\begin{center}
\includegraphics[width=0.85\linewidth]{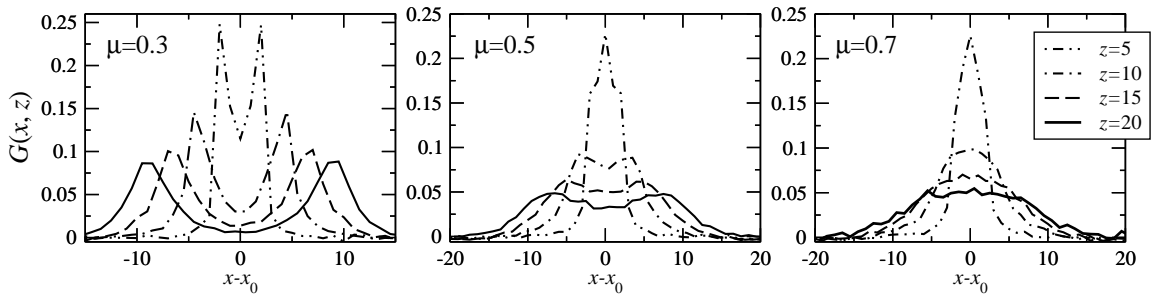}
\caption{Simulation results for the mean response $G(x,z)$ at
different depths $z$ in the frictional case. The value of the applied
force is 
$F=3$ (in units of $p$).
Each graph corresponds to a different value of the
coefficient of friction $\mu$. The total number of layers is $L=20$.\label{fig:profiles_i}} 
\end{center}
\end{figure*}

{\it Hexagonal packing of frictionless grains---} Consider a hexagonal
packing  of massless and  frictionless disks  confined by  an external
pressure     applied     on    the     walls     of    the     packing
(c.f.~Fig.~\ref{fig:packing}).                 In                Refs.
\cite{ostojic04:2,ostojic05-1},  we examined  this packing  under zero
vertical and  large horizontal pressure.  While  this situation allows
an instructive comparison between the force ensemble and the $q$-model
\cite{coppersmith96},   physically  it   corresponds  to   a  singular
limit, where there is no relevant scale for the applied overload \cite{previous}. Here we study the  practically relevant case where the pressure
is isotropic.  We have  obtained similar results (not shown here) for
grains confined by gravity instead of vertical pressure.

In  the absence  of friction,  the contact  forces are  normal  to the
grains.   As  each  grain  shares  six contacts  with  its  neighbors
[c.f.~Fig.~\ref{fig:packing}~(b)], there  are three independent forces
and two equations  of force balance per grain.  The force ensemble can
thus be  parametrized by specifying  one force per grain,  for example
$N_4$.  In this  case, the dimension of ${\cal  E}$ is $L(2L+1)$, and
its boundaris are determined by
the requirement that all the forces in the packing be repulsive. The
sampling was implementing via a ``hit and run'' algorithm \cite{vempala05}.

The  results  obtained  from  Monte  Carlo simulations  are  shown  in
Fig.~\ref{fig:pressure}.     Fig.~\ref{fig:pressure}(a)    shows   the
response  at  various  depths  to  an overload  $F=10$  (in  units  of
pressure), for several system sizes.  Close to the top of the packing,
$G(i,j)$  displays  two  sharp  peaks  along  the  lattice  directions
emanating from  the point of  application of the overload.   Deeper in
the packing, the  two peaks broaden, and an  increasing amount of load
is transfered towards the  center.  Progressively the response becomes
flatter,   and   at   a   given   depth  the   two   peaks   disappear
completely. Below that depth, only  a broad central peak persists, and
its width increases linearly with depth. These results are independent
of the size of the simulated system.

The dependence  on the magnitude of  the overload $F$  is displayed in
Fig.~\ref{fig:pressure}(b),  where the  response is  shown at  a depth
$i=15$.   For   small  overloads,  $G(i,j)$  is   single  peaked,  and
essentially independent of $F$, which shows that the variations of the
forces  are linear  in  applied  overload. As  $F$  is increased,  the
response becomes flatter and progressively two peaks appear. For large
overloads,  these peaks  are located  on the  lattice  directions.  In
conclusion,  for   increasing  overloads,  the   crossover  between  a
double-peaked and  a single-peaked response  is shifted deeper  in the
packing.

\begin{figure*}
\begin{center}
\includegraphics[width=0.85\linewidth]{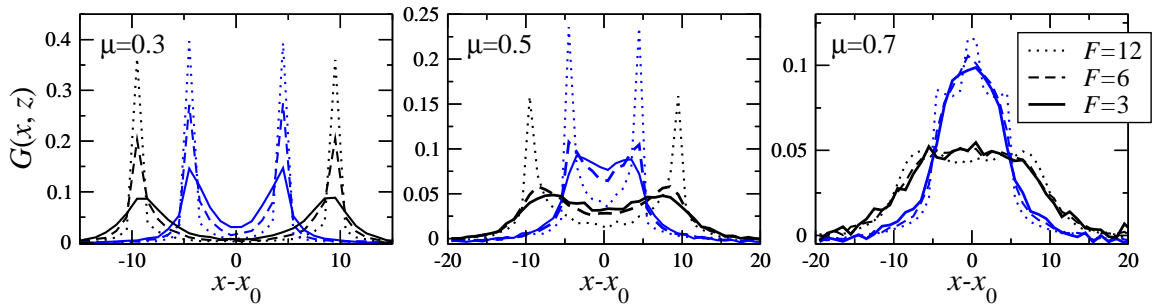}
\caption{The mean response $G(x,z)$ for different values of the
overload $F$ 
(in units of $p$) for 
the frictional case. Data are shown for two different
layers: $z=20$ (black), and $z=10$ 
(blue).
Each graph corresponds to a
different value of 
$\mu$. The total number of layers is $L=20$. \label{fig:profiles_f}}
\end{center}
\end{figure*}
{\it Rectangular  packing of frictional grains---} In  the presence of
friction,  each  contact force  possesses  a  tangential component  in
addition to  the normal one.   Moreover, torque balance on  each grain
must  be  taken into  account.   This  implies  that there  are  three
unknowns per  grain in a hexagonal  packing, a situation  for which we
found  reasonably-sized  systems impossible  to  simulate without  the
recourse  of  impractically  long  computer  runtimes.   We  therefore
considered a  rectangular packing, in  which grains in the  same layer
are  not  in  contact  (c.f.~Fig.~\ref{fig:packing}~(c)),  a  geometry
previously studied  in \cite{breton02},  but in a  different framework
\cite{differ}.   In  the  absence  of  friction,  the  response  of  a
rectangular packing is trivial,  as the overload propagates exclusively
along the lattice directions.  This setting thus allows to isolate the
effects of friction, which is the {\it only\/} source of disorder.

On each grain, there are  now four unknown forces and three equations,
so that the force ensemble can again be parametrized by specifying one
force per grain, which we choose to be $T_4$. The boundaries of ${\cal
E}$
, whose dimension is again $L(2L+1)$,
  are determined  by  the  requirement that  all  normal forces  be
repulsive and  that all tangential forces  satisfy Coulomb's criterion
$|T_i|\leq\mu N_i$, where $\mu$ is the coefficient of friction. Due to
this second requirement, the shape  of ${\cal E}$ is  complicated,
and we use a ``ball-walk'' sampling \cite{vempala05}.

\begin{figure}
\begin{center}
\includegraphics[width=0.8\linewidth]{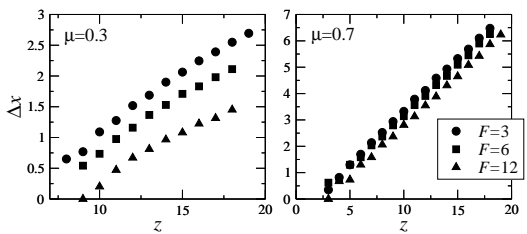}
\caption{Broadening of the 
response, 
quantified by
$\Delta x(z)= \left(\int x^2 G(x,z) dx-\left[\int x G(x,z) dx\right]^2\right)^{1/2}$.
 For $\mu=0.3$, the
sum was restricted to $x>x_0$.
\label{fig:broadening}} 
\end{center}
\end{figure}

The  results of  the Monte  Carlo simulations,  with $F$  expressed in
units   of  $p$  are   displayed  in   Figs.~\ref{fig:profiles_i}  and
\ref{fig:profiles_f}.   In Fig.~\ref{fig:profiles_i}, the  response to
an  overload  $F=3$  is  shown  at  different  depths,  for  different
coefficients of friction $\mu$. For small $\mu$, the response consists
of two peaks along the lattice directions.  For $\mu=0.3$, these peaks
broaden         essentially         linearly        with         depth
(cf.~Fig.~\ref{fig:broadening}),   in   agreement  with   experimental
findings \cite{geng03}.  For $\mu=0.5$,  the peaks are very broad, and
merge  into a  single one  below a  given depth.   For  $\mu=0.7$, the
response consists  exclusively of a single, central  peak, whose width
broadens   linearly    with   depth   (cf.~Fig.~\ref{fig:broadening}).
Altogether,  this results  indicate that  the depth  of  the crossover
between  double and single-peaked  response decreases  with increasing
$\mu$.

The dependence of the response on the magnitude of the overload $F$ is
displayed in  Fig.~\ref{fig:profiles_f} for different  coefficients of
friction $\mu$.  Clearly, as  $F$ is increased,  the two  peaks become
more  prominent,  and propagate  deeper  in  the  packing, as  in  the
frictionless  case. For small  $F$, in  the single-peaked  regime, the
response is independent of the overload, in other words it is linear.

{\it  Discussion---}  The force  ensemble  yields remarkably  accurate
predictions  for  the mechanical  response  of  ordered packings.   In
particular, our results  for the dependence of the  crossover depth on
applied overload and friction are in all points comparable to those of
molecular dynamics simulations \cite{goldenberg05}, which describe the
grains  at a much  greater level  of detail.  The force  ensemble thus
provides a simple theoretical framework which predicts both the linear
response  described  by  elasticity  theory,  and  the  experimentally
relevant non-linear behavior.

A closer  look at the horizontal  forces on the  grains indicates that
the  reason  for this  cross-over  can  be  intuitively understood  by
comparing  the typical magnitudes  of $W_{ij}$  with the  typical size
$\delta$ of fluctuations  in the horizontal forces on  the grains.  In
our formulation,  these fluctuations are parametrized  by $N_4-N_3$ in
the frictionless case,  and $T_4$ in the frictional  case. Their sizes
are determined by  the requirement that normal forces  be positive and
that tangential ones obey Coulomb's criterion, so that $\delta$
increases  with $p$  and  $\mu$. For  $W_{ij}\gg  \delta$, the  packing
effectively becomes rectangular  (although horizontal contacts are not
broken),  and  the  overload   propagates  mainly  along  the  lattice
directions.     The     scenario     is    indeed     confirmed     by
Fig.  ~\ref{fig:pressure}(b) ---  increasing $F/p$  implies increasing
$W_{ij}/p$, which  in turn  makes the peaks  in the  response function
penetrate deeper into the packing. 
To complement this observation, we also find 
that the cross-over occurs  
at a layer $i$ where $W_{ij}\sim\delta_{ij}$.

We  believe that  it is  possible  to use  our findings  in the  above
paragraphs  to   obtain  a  field-theory  type   formulation  for  the
mechanical  response  of  hard  grain  packings,  and  it  remains  an
important future direction of our work.

We thank  
G. Barkema for his help with the Monte Carlo simulations on frictional
grains.    SO  is   financially  supported   by  the   Dutch  research
organization FOM (Fundamenteel Onderzoek der Materie).


\begin{thebibliography}{26}
\bibitem{previous} As a consequence, the response is independant of
  the overload, and self-similar with the system size.
\bibitem{differ} Our  results differ significantly from  those of Ref.
\cite{breton02}, wherein $T_4$ was chosen independently on each grain,
in  the  spirit  of  the  $q$-model  \cite{coppersmith96}.   In  Refs.
\cite{ostojic04:2,ostojic05-1},   we   showed   that  the   stochastic
prescription  of   the  $q$-model  on  a   hexagonal  packing  differs
significantly from that of the  force network ensemble. It is thus not
a surprise to find a distinction between the two also in a rectangular
packing of frictional grains.
\end{thebibliography}
\end{document}